\documentclass[useAMS,usenatbib]{mn2e}

\usepackage{bm}
\usepackage{amssymb}
\usepackage{natbib}
\usepackage{graphicx}
\usepackage{psfrag}
\usepackage{pst-grad}

\newcommand{\mnras}{MNRAS}
\newcommand{\apj}{ApJ}
\newcommand{\apjl}{ApJ}{

\newcommand{\aap}{A\&A}

\newcommand{\ssr}{Space Science Rev.}
\newcommand{\prd}{Phys. Rev. D}
\newcommand{\jcap}{JCAP}

\newcommand{\eqb}{\begin{eqnarray}}
\newcommand{\eqe}{\end{eqnarray}}
\newcommand{\eqbn}{\begin{eqnarray*}}
\newcommand{\eqen}{\end{eqnarray*}}
\newcommand{\diff}{\textrm{d}}

\newcommand{\pdiff}[2]{\frac{\partial #1}{\partial #2}}

\title[Fermi acceleration at ultra-relativistic shocks]{On the maximum energy of shock-accelerated cosmic rays at ultra-relativistic shocks}
\author[B. Reville and A.~R. Bell]{B. Reville$^{1}$\thanks{E-mail:
b.reville@qub.ac.uk}, A.~R. Bell$^2$ \\
$^1$Centre for Plasma Physics, Queen's University Belfast, University Road, Belfast BT7 1NN\\
$^2$Clarendon Laboratory, University of Oxford, Parks Road, Oxford OX1 3PU}

\begin{document}

\date{Accepted and Received \dots}

\pagerange{\pageref{firstpage}--\pageref{lastpage}} \pubyear{2014}

\maketitle

\label{firstpage}

\begin{abstract}
The maximum energy to which cosmic rays can be accelerated at weakly-magnetised ultra-relativistic shocks is investigated.
We demonstrate that for such shocks, in which the scattering of energetic particles is mediated exclusively by ion skin-depth scale structures,
as might be expected for a Weibel-mediated shock, there is an intrinsic limit on the maximum energy
to which particles can be accelerated. This maximum energy is determined from the requirement that particles must be isotropised 
in the downstream plasma frame before the mean field transports them far downstream, and falls considerably short of what is required to 
produce ultra-high-energy cosmic rays. To circumvent this limit, a highly disorganised field is required on 
larger scales. The growth of cosmic-ray induced instabilities on wavelengths much longer than the ion-plasma skin depth, 
both upstream and downstream of the shock, is considered. While these instabilities may play an 
important role in magnetic field amplification at relativistic shocks, on scales comparable to the gyroradius of the most energetic
particles, the calculated growth-rates have insufficient time to modify the scattering. Since strong modification is a necessary condition for particles in the
downstream region to re-cross the shock, in the absence of an alternative scattering mechanism, these results imply that 
acceleration to higher energies is ruled out. If weakly magnetised ultra-relativistic shocks are disfavoured as high-energy particle
accelerators in general, the search for potential sources of ultra-high-energy cosmic rays can be narrowed.
\end{abstract}


\section{Introduction}

Ultra-relativistic shocks are known to occur in the outflows of gamma-ray bursts (GRBs),
pulsar winds, and active-galactic nuclei (AGN). These non-linear structures are frequently
observed to be strong sources of non-thermal radiation, resulting from the inverse-Compton and
synchrotron emission of recently accelerated relativistic particles in the local photon and magnetic fields. 
Gamma-rays and neutrinos produced in hadronic interactions are also expected, although their 
radiative signatures are more difficult to detect. Ultra-relativistic shocks have also been suggested as potential sources 
of ultra-high energy cosmic rays  \cite[UHECRs, see][and references therein]{Bykovetal}, where the relativistic Fermi shock-acceleration  
mechanism is thought to be the primary mechanism for accelerating these particles to energies in excess of $10^{19}$~eV.

The theory of Fermi acceleration at relativistic shocks dates back over 30 years \citep{Peacock},
and is undoubtedly the most established mechanism for converting the bulk kinetic energy of a relativistic 
outflow, into non-thermal high-energy particles \cite[see][for a review]{KirkDuffy}.
The large Lorentz boosts required to transform between the upstream and downstream rest-frames, preclude the usual 
assumption of near-isotropy, which is frequently used in non-relativistic treatments. In this situation, the details of particle scattering on in situ electromagnetic 
fluctuations, plays a more important role \cite[see][for some examples of studies into 
the effects]{KirkSchneider87,LemoineRevenu06,NiemiecOstrowski}.
Considerable progress has followed recent particle-in-cell (PIC) simulations, which have succeeded in 
demonstrating from first principles, that relativistic shocks can accelerate particles through repeated scattering across a shock
\cite[e.g.][]{SironiSpitkovsky11,Stockemetal}. The particle scattering in these simulations is dominated by deflections in 
short wavelength, plasma skin-depth scale fluctuations ($\lambda \gtrsim c/\omega_{\rm pp}$, where 
$\omega_{\rm pp}=\sqrt{4\pi n e^2/\bar{\gamma}m_{\rm p}}$ is the relativistic 
plasma frequency, with $\bar{\gamma}$ the mean thermal Lorentz factor), 
driven by Weibel/two-stream like instabilities \cite[see][for a thorough numerical investigation]{sironietal13}.
However, if the gyroradius of the particles greatly exceeds the size of these skin-depth scale structures, 
in the absence of larger scale ($\lambda \gg c/\omega_{\rm pp}$) fluctuations, the particles
are tied to the mean field \citep{casseetal02,revilleetal08}, and subsequent crossings from downstream to upstream will 
be suppressed. Since the mean field, for most realisations, lies in the plane of the shock downstream, 
as pointed out by \cite{Achterbergetal}, unless cross 
field diffusion is close to the Bohm limit in the downstream region, a particle in the downstream has no chance of 
overtaking the shock. This requires the field to be dis-organised on scales close to the particles' Larmor radii in the downstream 
region. Thus, the ability to scatter high energy particles \emph{downstream} of the shock, is the determining factor with regard to
maximum energy. 

The most obvious recourse for producing fluctuations on the required scales, is to seed them in the upstream via streaming 
instabilities, initiated by the returning accelerated particles. While several such investigations have been carried
out \cite[e.g.][]{Revilleetal06, MilosNakar06, MedvedevZak}, the issue of long-wavelength perturbations relevant to 
UHECR scattering has not been satisfactorily addressed.

Here we determine the growth of long-wavelength plasma instabilities $(ck\ll \omega_{\rm pp})$ in the limit of ultra-relativistic shock velocities. In 
particular, we investigate instabilities occurring in shock-precursors, in which the energy density of the returning particle flux,
dominates the energy density in the upstream ion rest-frame, which is most likely the case for ultra-relativistic shocks. 
As we show, the ratio of these energy densities, in the upstream thermal ion frame, is of the order of $e_{\rm cr}/e_{\rm th}\sim \eta \Gamma_{\rm sh}^4$,
where $\eta$ is a measure of the efficiency with which the incoming energy density is reflected as accelerated protons. 
Current simulations indicate that this fraction is of the order of $\eta \sim 1-10\%$
independent of shock Lorentz factor and magnetisation, \cite[below a critical magnetisation $\sigma<\sigma_{\rm crit}\sim10^{-3}$, ][]{sironietal13}.
For shocks with bulk Lorentz factor  $\Gamma_{\rm sh} > 10$, the cosmic-ray energy density clearly dominates. 
This modifies the general plasma conditions in the precursor of a shock propagating into an ambient electron-proton plasma,
and must be taken into account when investigating plasma instabilities in the upstream pre-cursors of relativistic shocks.

The outline of the paper is as follows. In the next section, we discuss the scattering and maximum energy expected in the 
small-angle scattering limit when the mean field component is included. The bulk properties and conditions in the shock-precursor 
are described in Section \ref{Sec:Param}. In section \ref{Sec:instability} 
we use these conditions to determine the dispersion relation for long-wavelength modes $\lambda \gg c/\omega_{\rm pp}$. Section \ref{Sec:acc} applies these results,
in the context of cosmic-ray acceleration. The growth of plasma instabilities in the downstream region is also determined. We conclude with some additional discussion 
on the implications of these results.

\section{Particle scattering and maximum energy in small-scale turbulence}

We focus on  ultra-relativistic ($\Gamma_{\rm sh}\gg1$) electron-ion shocks, as potential sources of cosmic-ray protons and nuclei. It is convenient when 
talking about such shocks to introduce the magnetisation parameters in the upstream and downstream regions
\eqb
~~~~~~~~\sigma_{\rm u} = \frac{B_{\rm u}^2}{4\pi n_{\rm u} m_{\rm p} c^2}~,~~~~~~ \sigma_{\rm d} = \frac{B_{\rm d}^2}{4\pi \bar{\gamma} n_{\rm d} m_{\rm p} c^2}\enspace ,
\eqe
where all quantities here are measured in the local plasma frame:
$B_{\rm u,d}^2/4\pi$ represents the total magnetic pressure in each region, and $\bar{\gamma}\approx \Gamma_{\rm sh}$ is the mean 
thermal Lorentz factor of the downstream plasma \cite[e.g.][]{BlandfordMcKee76}. We consider the most common ultra-relativistic shock scenario, 
in which the shock propagates into a plasma with $w_{\rm u} =e_{\rm u}+p_{\rm u}\approx n_{\rm u} m_{\rm p} c^2$. 
Here $w_{\rm u}$ is the specific enthalpy of the upstream plasma and $p_{\rm u}$ and $e_{\rm u}$ 
are the upstream pressure and energy density respectively.  
Considering for the moment the mean fields alone, as expected from the ultra-relativistic shock jump conditions, 
the magnetisation downstream is: a) for parallel shocks, $B_{\rm d}=B_{\rm u}$ and $\sigma_{\rm d}\ll\sigma_{\rm u}$, or b) for perpendicular shocks,  
$B_{\rm d}/B_{\rm u} = n_{\rm d}/n_{\rm u} \approx \bar{\gamma}$ such that  $\sigma_{\rm d}\sim \sigma_{\rm u}$, with oblique shocks falling in 
between these two extremes \cite[see][for a review of relativistic MHD jump conditions]{KirkDuffy}. 

Particle-in-cell simulations of weakly-magnetised ($\sigma_{\rm u}<10^{-3}$) relativistic shocks, have demonstrated that non-thermal particle acceleration,
is a natural outcome of the shock formation process.
For such shocks, the energy dissipation of the upstream plasma is mediated by Weibel or filamentation-type instabilities, resulting in ion skin-depth scale structures that thermalise the 
incoming flow \citep{SironiSpitkovsky11, Haugboll}. The self-generated fields, at least on the scale of current simulations, show 
$\sigma_{\rm d} \gg \sigma_{\rm u}$, with typical values of the order $\sigma_{\rm d}\sim 10^{-2}$, in contrast to the MHD calculations made
above. 

The scattering of particles on these fluctuations also results in a small fraction, of the order of $1 \%$, escaping back into the upstream, 
thus providing the injection mechanism in the Fermi acceleration process \citep{sironietal13}. These particles can scatter further on the Weibel generated 
structures, and cross the shock multiple times. We consider the relevant energy and length scales involved.

It is assumed that particles have zero probability of escaping infinitely far ahead of the shock, and that transport downstream 
is the only escape channel (although see below for the implications of this). Hence, in the absence of radiative losses, or finite time limitations, 
the maximum energy is ultimately determined by the ability to scatter particles \emph{downstream} of the shock.
For an ultra-relativistic shock moving along the $z$-axis with speed $\beta_{\rm sh}$, any particles capable of overtaking the shock must have their 
parallel velocity component $\beta_{z} > \beta_{\rm sh} \approx 1/3$. Since $\beta_{z} = \beta\cos\theta \approx \cos\theta$, 
on transforming the pitch angle into the upstream rest frame, it follows that all particles are
confined to a narrow cone of half-opening angle $\theta\sim 1/\Gamma_{\rm sh}$.
Once they leave this cone, they are quickly over-taken by the shock. If the pitch angle, $\mu\equiv \cos\theta$, is still closely aligned with
the shock normal, ($|\mu| \lesssim \beta_{\rm sh}$), as is typically the case, a Lorentz transformation to the downstream frame results in a 
reduction of the cosmic-ray energy $E'_{\rm cr}= \Gamma_{\rm rel} E_{\rm cr}(1-\beta_{\rm rel}\beta_{\rm cr}\mu)\sim E_{\rm cr}/\Gamma_{\rm sh}$
where $\Gamma_{\rm rel}\approx\Gamma_{\rm sh}/\sqrt{2}$ is the Lorentz factor of the downstream flow measured by an upstream observer. 
For simplicity, in what follows, we will assume that this is always the case neglecting factors of order unity, 
i.e. a particle with Lorentz factor $\gamma$ in the upstream has $\gamma_{\rm d} = \gamma/\Gamma_{\rm sh}$
in the downstream, and vice-versa.

For $\sigma_{\rm d}\gg \sigma_{\rm u}$, the downstream particle gyro radius in the self-generated field matches the scattering structure scale $\lambda_{\rm d}$ 
when 
\eqb
\gamma_{\rm d} = \frac{eB_{\rm d} \lambda_{\rm d}}{ m_{\rm p} c^2} = \bar{\gamma}\sigma_{\rm d}^{1/2}\frac{\lambda_{\rm d}}{c/\omega_{\rm pp}} \enspace .
\eqe
Particles with Lorentz factor exceeding this value in the downstream frame enter the small-angle scattering regime, which allows them to 
diffuse back across the shock. Recalling that simulations suggest $\sigma_{\rm d} \sim 10^{-2}$ and $\lambda \sim 10-20~c/\omega_{\rm pp}$, 
it is clear that even mildly supra-thermal particles have sufficient energy to enter this scattering regime, as is indeed observed in PIC simulations. 
However, as the particles are accelerated to much larger energies, these small-angle deflections ultimately become insignificant, 
and the mean field must again be considered.  Unless some additional scattering field exists at longer wavelengths, particles
will simply gyrate in the large-scale mean fields, which recedes at a velocity $\approx c/3$ in the downstream. 
Cross field diffusion close to the Bohm limit in the downstream region is required for a particle to have any chance of returning to the shock \citep{Achterbergetal}. 
This requires fluctuations to be generated at, or close to, the scale of particles' Larmor radii in the downstream region.

Following \cite{KirkReville}, we quantify the small angle scattering behaviour in terms of an angular diffusion coefficient, $D_\theta =\langle\Delta\theta^2\rangle/{2\tau}$
where $\tau$ is the mean time between scatterings, and $\Delta\theta$ the average deflection angle. For ultra-relativistic particles, the scattering time 
is approximately the light-crossing time of the Weibel-generated structures. 
The average deflection angle at each scattering, for a particle with Lorentz factor $\gamma_{\rm d}$, 
in the downstream region is 
\eqb
\Delta \theta_{\rm d}({\gamma}_{\rm d}) \approx \frac{e \sqrt{\delta B_{\rm d}^2} \lambda_{\rm d}}{\gamma_{\rm d}m_{\rm p} c^2} =
\frac{\bar{\gamma}}{\gamma_{\rm d}}\frac{\lambda_{\rm d}}{c/\omega_{\rm pp}} \sigma_{\rm d}^{1/2} \enspace,
\eqe
from which we can evaluate
\eqb
\label{Eq:Dtheta}
D_\theta \approx   \left(\frac{\bar{\gamma}}{\gamma_{\rm d}}\right)^2\frac{\lambda_{\rm d}}{c/\omega_{\rm pp}} \sigma_{\rm d}\omega_{\rm pp}\enspace.
\eqe 
So far, we have neglected the effect of the mean field in the downstream. The mean isotropisation time for a distribution of particles is $\sim D_\theta^{-1}$, 
which becomes increasingly large for $\gamma_{\rm d}\gg\bar{\gamma}$. Above a critical energy the particles will again 
return to quasi-helical orbits. This can be expected to occur when the isotropisation time exceeds the Larmor period in the 
shock compressed mean field, e.g. \cite{LemoinePelletier10}
\eqb
D_\theta \Omega_{\rm g}^{-1} = D_\theta \frac{\gamma_{\rm d} m_{\rm p} c}{e \langle B_{\rm d}\rangle} < 1\enspace .
\eqe
Hence, in the absence of larger scale fluctuations close to spatial resonance with cosmic-ray Larmor radius in the downstream region,
the maximum Lorentz factor is limited to
\eqb
\label{Eq:gammaxd}
\gamma_{\rm d, max} \sim \bar{\gamma} \frac{\lambda_{\rm d}}{c/\omega_{\rm pp}} \sigma_{\rm d}\sigma_{\rm u}^{-1/2}
\eqe
as measured in the downstream frame, with $\sigma_{\rm u}$ the magnetisation far upstream. \cite{LemoinePelletier10} arrive at the same limiting energy.
If particles at this energy can somehow escape upstream, they will gain another 
factor $\bar{\gamma}$, suggesting a maximum energy 
\eqb 
\label{Eq:EMAXXX}
E_{\rm max} \approx 
\left(\frac{{\Gamma_{\rm sh}}}{100}\right)^2 \left(\frac{\lambda_{\rm d}}{10c/\omega_{\rm pp}}\right)
\left(\frac{\sigma_{\rm d}}{10^{-2}}\right)\left(\frac{\sigma_{\rm u}}{10^{-8}}\right)^{-1/2}~\mbox{PeV~,} 
\eqe
where we have chosen parameters that might be relevant for an external GRB shock. In what follows, we investigate the 
possibility for generating magnetic fluctuations on sufficient scale and amplitude to facilitate acceleration to higher energies than
the above limit. Finally, we note that a distant observer will measure the downstream energy, boosted by the relative Lorentz factor 
$\gamma_{\rm obs}= \Gamma_{\rm rel}\gamma_{\rm d}$. However, for the sources of interest, by the time the particles are released
into the interstellar or intergalactic medium (ISM/IGM), the flow will have decelerated appreciably, in which case the maximum energy will be reduced via adiabatic losses by a factor 
$\sim \bar{\gamma}$. Hence particle escape is an important factor, for accelerating to high energies.

\section{Reflected particles and electron drift}
\label{Sec:Param}

As mentioned in the previous section, using the MHD shock jump conditions, the mean thermal energy per particle,
resulting from the thermalisation of a cold incoming fluid, as measured in
the downstream plasma frame is $\bar{\gamma}\approx\Gamma_{\rm sh}$. It is assumed given that a small fraction
of these shocked particles can escape back into the upstream. The mean energy for such particles, now measured in the upstream rest frame, 
is $\gamma_{\rm u}\approx\Gamma_{\rm sh}^2$.
For simplicity, we assume that the particle mean-free path is longer than the shock transition layer, which need not, and for low-energy cosmic rays
in a Weibel-mediated shock, most likely is not, the case. Any modification however, is expected to be small, and decreases with increasing particle energy.
As mentioned in the previous section, a Lorentz transformation from upstream to downstream results, on average, in a reduction of a cosmic-ray's energy,
although the over-all energy gain per cycle, is still a net gain. The particles energy is of course approximately constant as measured by a distant observer, 
however, if the particle is adiabatically coupled to the downstream flow, it will lose energy as the flow decelerates before finally releasing particles. 
With regards UHECR acceleration, this can have a significant effect on the maximum energy that can be 
achieved at a shock, since the initial $\Gamma^2$ increase is effectively wasted unless the particle can ultimately escape to infinity, 
upstream of the shock, before it has decelerated appreciably. 

Here, the possible contribution of relativistic shocks to the cosmic-ray production is addressed.
We focus, initially, on the first generation of shock reflected particles, i.e. those that, as measured 
in the upstream rest frame, have Lorentz factor $\Gamma_{\rm cr}\approx \Gamma_{\rm sh}^2$,
and for simplicity consider a pure electron-proton shock. 
In the shock rest frame, the far upstream plasma can be treated as cold beam, with energy 
density $\approx \Gamma_{\rm sh}\bar{n}_{\rm p} m_{\rm p} c^2$, where $\bar{n}_{\rm p}$ 
is the ambient proton density measured in the shock frame. If a fraction $\eta$ of this energy 
leaks back into the upstream\footnote{
The incident energy flux of the upstream thermal plasma, measured in the shock frame 
is ${T}^{01}_{\rm in} = \Gamma_{\rm sh}^2 \beta_{\rm sh} n_0m_{\rm p} c^2$, where $n_0$ is the upstream proper density.
We define $\eta$ as the fraction of this energy flux that is reflected ${T}^{01}_{\rm ref} = -\eta \Gamma_{\rm sh}^2 \beta_{\rm sh} n_0m c^2$.
}  with average velocity $-v_{\rm sh}$ in the shock frame, 
on transforming back to the upstream proton frame, this 
reflected component has energy density $\approx 4\eta \Gamma_{\rm sh}^4n_{\rm p} m_{\rm p} c^2$,
i.e. if $\Gamma_{\rm cr}\approx \Gamma_{\rm sh}^2$, the cosmic-ray density, as measured in the proton frame is
$n_{\rm cr}\approx4\eta\Gamma_{\rm sh}^2n_{\rm p}$. If, as simulations suggest, $\eta\sim 1\%$, for $\Gamma_{\rm sh}\gg1$,  
the number density of the cosmic rays can easily 
exceed that of the background in this frame. As we demonstrate, this can considerably alter the equilibrium 
plasma conditions in the upstream. This effect, is not confined to the lowest energy particles. The efficiency factor for returning cosmic rays
is the sum over all energetic particles $\eta \propto \int_{\gamma_{min}}^\infty n(\gamma)d\gamma$.
Far upstream, where only higher energy particles can reach, $\gamma_{\rm min}$ will increase, and unless $n(\gamma)$ is very flat, 
$\eta\Gamma_{\rm sh}^2$ will decrease upstream. Theory and simulations suggest $n(\gamma) \propto \gamma^{-2.2}$  
\cite[e.g.][]{Kirketal00,Achterbergetal, sironietal13}, such that $\eta$ will decrease slightly faster than $\gamma_{\rm min}^{-1}$, 
Hence, beyond a critical distance upstream $\eta \Gamma_{\rm sh}^2 < 1$, and the current density is too low to have a
significant effect. In what follows, unless otherwise stated, we will take $\eta\Gamma_{\rm sh}^2\gg 1$ to hold at all times.
 
We recall that the conditions of charge and current neutrality are,
\eqb
n_{\rm e} = n_{\rm p}+n_{\rm cr} \mbox{~~and~~} n_{\rm e}\beta_{\rm e} = n_{\rm p}\beta_{\rm p}+n_{\rm cr}\beta_{\rm cr}
\eqe
which, provided all terms are appropriately defined, is frame independent. Considering initially, the ambient protons' rest frame, 
the electrons provide a return current to satisfy the above conditions. 
Using the above value for $n_{\rm cr}$, the electrons will drift with velocity 
\eqb
\label{Eq:betae}
\beta_{\rm e} = \frac{4\eta\Gamma_{\rm sh}^2}{1+4\eta\Gamma_{\rm sh}^2}\beta_{\rm cr}
\eqe
with respect to the protons, to cancel the cosmic-ray current.

Since the electric field approximately vanishes in the electrons' rest-frame, it is advantageous to work in this frame. 
To prevent confusion, we use upper case $\Gamma$ for quantities measured in the upstream proton frame, and lower case 
$\gamma$ for quantities measured in electron frame.
From (\ref{Eq:betae}) it follows that the protons, by symmetry, now drift with respect to the electrons with Lorentz factor  
\eqb
\label{Eq:Gammae}
\gamma_{\rm p}  \approx
\sqrt{2\eta}\Gamma_{\rm sh}\enspace, 
\eqe 
while the Lorentz factor of the shock relative to the electrons is significantly reduced:
\eqb
\gamma_{\rm sh} \approx \frac{1}{\sqrt{8\eta}}\enspace .
\eqe 
The cosmic rays, which had Lorentz factor $\Gamma_{\rm cr}$ in the proton frame, now have  
\eqb
\label{Eq:Gammacr}
\gamma_{\rm cr} \approx\frac{\Gamma_{\rm cr}}{\sqrt{8\eta}\Gamma_{\rm sh}}\enspace .
\eqe 
Since both $\gamma_{\rm p},{\gamma}_{\rm cr}\gg 1$ under our assumption, 
it follows from the charge and current conditions, that $n_{\rm e} \approx 2n_{\rm p} \approx 2n_{\rm cr}$
in the electron rest frame.

It is also necessary to consider the equilibrium configuration of the magnetic field in this limit. 
Since, to lowest order, the magnetic field is expected to be frozen into the 
electrons, the net drift of electrons with respect to the protons can have a significant impact, even without 
resorting to plasma instabilities. Sufficiently far upstream, beyond the range
of influence of the cosmic rays, there is no drift between the protons and electrons, and the 
electric field is approximately zero in the plasma rest-frame. As the shock approaches, the cosmic-rays 
cause the electrons to drift with respect to protons. From charge conservation
\eqb 
\pdiff{n_{\rm e}}{t}+\bm{\nabla}\cdot(n_{\rm e} \bm{v}_{\rm e})=0\enspace ,
\eqe 
and the magnetic induction equation,
\eqb
\label{Eq:Induction}
\pdiff{\bm{B}}{t} = -c \bm{\nabla}\times\bm{E} \approx \bm{\nabla}\times(\bm{v}_{\rm e}\times\bm{B})\enspace ,
\eqe
for flow in one dimension, it follows that ${B_\bot}/{n_{\rm e}} = \mbox{const}$. Inside the shock precursor, 
in the rest frame of the electrons, the electric field remains zero, and the density is 
$n_{\rm e} = 2 n_{\rm p} = 2\sqrt{2\eta}\Gamma_{\rm sh}n_{\rm p,0}$, where $n_{\rm p,0}$ is the 
proton proper density, which is also the electron density far upstream. Hence, the perpendicular 
component of the magnetic field in the rest frame of the compressed electrons, is 
$B_\bot =  2\sqrt{2\eta}\Gamma_{\rm sh}B_{\bot}^0$. Note that in the shock frame, the fields are
still $\hat{B}_\bot = \Gamma_{\rm sh}B_{\bot}^0$, and $\hat{E}_\bot= -\bm{\beta}_{\rm sh}\times\hat{B}_\bot$, 
such that the downstream fields still satisfy the Rankine-Hugoniot relations for a single fluid weakly magnetised 
MHD shock \cite[e.g.][]{KennelCoroniti84}.

Finally we note that, as mentioned above, higher energy cosmic-rays extend further upstream from the 
shock than lower energy ones. Since the efficiency factor depends on all the cosmic-rays that contribute to the 
returning flux, i.e. $\eta$ is a function of $\gamma_{\rm min}$, it will also have a spatial dependence. 
The cosmic-ray number density gradually decreases upstream, eventually to the level where the field is unmodified. 
The above effect will produce a large-scale inhomogeneous magnetic field, with associated current 
$\bm{j}_\bot=\bm{\nabla}\times \bm{B_\bot}$. The precise details of this current will depend on the shape of the spectrum, 
as well as the scattering upstream. However, since $j_\bot \ll j_{\rm cr}$ we ignore it in the following. 

\section{Linear stability analysis of foreshock region}
\label{Sec:instability}

We continue to work in the electron drift frame.
Using the conditions described in the previous section, i.e. equations (\ref{Eq:Gammae}) -- (\ref{Eq:Gammacr}), 
we investigate the evolution of linear perturbations to the background plasma
$\bm{B}=\bm{B}_0+\bm{B}_1$ etc. This is 
not a straightforward task for relativistic shocks, since it is not immediately clear how to construct 
an equilibrium solution. As demonstrated in the previous section, unless the ambient field and shock normal
are aligned to within $1/\sqrt{8\eta}\Gamma_{\rm sh}$, the field will be highly oblique,
and a zeroth order $\bm{v}_0\times\bm{B}_0$ force will act on each species that drifts with respect to 
the electrons.

To make progress, we approximate the cosmic rays as infinitely rigid ($n_{\rm cr,1} = v_{\rm cr,1} = 0$), 
and the electrons as a massless fluid. The protons will still accelerate on account of the $\bm{v}_0\times\bm{B}_0$
force, and generate an oblique current, which will be neutralised by the massless electrons. Hence, the electron 
frame must itself accelerate. This effect can be accounted for with the inclusion of a gravity term in the proton 
equation of motion.

Both the electrons and protons satisfy charge conservation
\eqb
\label{Eq:ChrgCons}
\frac{\partial n_\alpha }{\partial t} + 
\bm{\nabla} \cdot (n_\alpha \bm{\beta}_\alpha c) = 0 
\eqe
where $\bm{\beta}_\alpha$ is the fluid velocity of each species in units of $c$,
and both species are treated as cold fluids
\eqb
\frac{\diff\bm{p}_\alpha }{\diff t} 
= {q_\alpha } \left(\bm{ E} +
 \bm{\beta}_\alpha  \times \bm{B}\right) + \gamma_\alpha m_{\alpha}\bm{g}
\eqe
The gravity term $\bm{g}$, representing the acceleration of the non-inertial electron frame, by definition 
acts to cancel the zeroth-order $\bm{\beta}_{\rm p,0}  \times \bm{B}_0$ force, should it exist.
Since the electrons are massless, $d\bm{p}_{\rm e}/dt=m_{\rm e}\bm{g} = 0$ and the electric field is 
$\bm{E}=-\bm{\beta}_{\rm e}\times\bm{B}_0$, where $\bm{\beta}_{\rm e}$ is a first order quantity.

Since ${\beta}_{\rm e} \ll 1$, we can safely neglect the displacement current in Ampere's law, 
\eqb
{\bf \nabla \times B} = \frac{1}{c}\frac{\partial {\bf E}}{\partial t}
+{4\pi}\sum_{\rm \alpha=p,e} n_\alpha q_\alpha \bm{\beta}_\alpha  \enspace ,
\eqe 
such that the electron velocity can be expressed in terms of the other perturbed quantities: 
\eqb
\bm{\beta}_{\rm e} = \frac{n_{\rm p0}}{n_{\rm e0}}\left[\bm{v}_1 + \frac{n_{\rm p1}}{n_{\rm p0}}\bm{v}_0 - \frac{c\Omega}{\omega^2_{\rm pp}}\bm{\nabla}\times\bm{b}_1\right]\enspace ,
\eqe
where $\bm{\beta}_{\rm p}=\bm{v}_0+\bm{v}_1$ is the proton velocity, $\Omega = eB_0/\gamma_0 m_{\rm p}c$ the relativistic proton gyro frequency, $\omega_{\rm pp} = (4\pi n_{\rm p}e^2/\gamma_0m_{\rm p})^{1/2}$ the proton plasma 
frequency, and $\bm{b}_1 = \bm{B}_1/B_0$ the normalised magnetic field fluctuations.
As shown in the previous section, in the ultra-relativistic limit, we have $n_{\rm e0}=2n_{\rm p0}=2n_{\rm cr}$, and the above becomes
\eqb
\bm{\beta}_{\rm e} = \frac{1}{2}\bm{v}_1 + \frac{1}{2}\frac{n_{\rm p1}}{n_{\rm p0}}\bm{v}_0 - \frac{1}{2}\frac{c\Omega}{\omega_{\rm pp}^2}\bm{\nabla}\times\bm{b}_1
\enspace .
\eqe

Together with equations (\ref{Eq:Induction}) and (\ref{Eq:ChrgCons}), it can be shown that 
\eqb
\label{Eq:LinB}
&&\frac{d}{dt}\left[
\frac{\partial\bm{b}_1}{\partial t}+\frac{c^2}{2}\frac{\Omega}{\omega^2_{\rm pp}} 
(\bm{b}_0\cdot\bm{\nabla}) \left(\bm{\nabla}\times \bm{b}_1\right)
\right]= \\
&&~~~~\frac{c}{2}\left[(\bm{b}_0\cdot\bm{\nabla})\left( \frac{d\bm{v}_1}{d t} - c \bm{v}_0(\bm{\nabla}\cdot\bm{v}_1)\right)
-\bm{b}_0\frac{\partial}{\partial t}(\bm{\nabla}\cdot\bm{v}_1)\right] ,\nonumber
\eqe
where the convective derivative is with respect to the zeroth order proton velocity $d/dt = \partial/\partial t + c\bm{v}_0\cdot \bm{\nabla}$.
Similarly, linearising the momentum equation produces 
\eqb
\label{Eq:LinMom}
&&\frac{\partial}{\partial t}\frac{d}{dt}\left[\bm{v}_1+\gamma_0^2(\bm{v}_1\cdot\bm{v}_0)\bm{v_0}\right]=\\
&&~~~~~\frac{\Omega}{2}\left[\frac{\partial \bm{v}_1}{\partial t}\times \bm{b}_0 + c(\bm{b}_0\cdot\bm{\nabla})\bm{v}_0\times\bm{v}_1\right] -\nonumber\\
&&~~~~~~~~~\frac{c}{2}\frac{\Omega^2}{\omega^2_{\rm pp}}\left[\bm{b}_0\times\left(\bm{\nabla}\times\frac{\partial \bm{b}_1}{\partial t}\right) \right. +\nonumber\\ 
&& ~~~~~~~~~~~~~~c(\bm{b}_0\cdot\bm{\nabla})\bm{v}_0\times(\bm{\nabla}\times\bm{b}_1)\Big]  ~.\nonumber
\eqe
While these equations are closed, and can be used to derive a complete dispersion tensor,
the result is cumbersome and is too complicated to proved the desired insight. However, since we are only concerned with long-wavelength 
fluctuations, we can simplify the problem considerably. For characteristic time scales $\omega^{-1}$ and length scales $k^{-1}$, 
the ordering of the terms on the right hand side of equation (\ref{Eq:LinMom}) is
$$
\frac{\omega}{\Omega} \enspace : \enspace \frac{kc}{\Omega}  \enspace : \enspace \frac{kc}{\omega_{\rm pp}}\frac{\omega}{\omega_{\rm pp}} \enspace : \enspace 
\frac{c^2k^2}{\omega_{\rm pp}^2} $$
such that for long-wavelength fluctuations ($\gg c/\omega_{\rm pp}$)
the final two terms on the right involving the fluctuating magnetic field can be neglected, and the 
remaining terms 
\eqb
\label{Eq:LinMom2}
&&\frac{\partial}{\partial t}\frac{d}{dt}\left[\bm{v}_1+\gamma_0^2(\bm{v}_1\cdot\bm{v}_0)\bm{v_0}\right]=\nonumber\\
&&~~~~~\frac{\Omega}{2}\left[\frac{\partial \bm{v}_1}{\partial t}\times \bm{b}_0 + c(\bm{b}_0\cdot\bm{\nabla})\bm{v}_0\times\bm{v}_1\right]
\eqe
contain only the different components of $\bm{v}_1$, which can be solved. 
We consider the dispersion of plane-waves satisfying equation (\ref{Eq:LinMom2}) for the
two limiting cases of exactly parallel and perpendicular shocks.

\subsection{Parallel shock}

Although parallel ultra-relativistic shocks, i.e. those for which the magnetic field is aligned with the shock normal to within 
$1/\Gamma_{\rm sh}$, are likely to be very rare in Nature, (assuming a random distribution of
shock propagation directions and ambient field orientations), they have nevertheless been, by far, the most commonly studied case  
\cite[e.g.][]{MilosNakar06,Revilleetal06}. We use our formalism to investigate the behaviour of long-wavelength 
linear fluctuations at such shocks. Parallel shocks represent a singular case for relativistic shocks, since 
reflected particles can in principle propagate upstream for large distances, provided the upstream medium is 
sufficiently uniform. Similar to supernova remnant shocks, the particles are expected to be self confined due to 
self-generated turbulence \cite[e.g.][]{Belletal13}.

We look for plane wave solutions 
to equation (\ref{Eq:LinMom2}), $\bm{v}_1 = \bar{\bm{v}}{\rm exp}[{\rm i}(\bm{k}\cdot\bm{x} - \omega t)]$. 
Taking the shock normal in the negative $z$ direction, i.e. $v_0>0$, the dispersion relation reads
\eqb
\omega^2(\omega-ck_zv_0)^2 - \frac{\Omega^2}{4}(\omega+ck_zv_0)^2 = 0
\eqe
which, provided $\bm{k}\cdot\bm{B}_0 \neq 0$, has always one unstable mode. The instability results from the 
uncompensated currents associated with the thermal background, which is maximised for perturbations 
along the mean field, i.e. $k=k_z$. The frequency for the growing mode is
\eqb
\label{Eq:OmegaPll}
\omega = \frac{1}{2}\left({ck_zv_0}-\frac{\Omega}{2}\right)+\sqrt{\left(\frac{ck_zv_0}{2}-\frac{\Omega}{4}\right)^2
-\frac{\Omega ck_zv_0}{2}}
\eqe
which has maximum growth rate and corresponding wavenumber  
\eqb
\label{Eq:PllMax}
\omega_{\rm max} = \left(\frac{1}{2} + \frac{1}{\sqrt{2}}{\rm i}\right)\Omega \mbox{  and  } k_{\rm max} =\frac{3}{2}\frac{\Omega}{cv_0}
\eqe
Thus the maximum growth occurs on scales close to the gyro-radius of the background drifting protons, which is 
smaller than that of the cosmic-rays by a factor $\sim \Gamma_{\rm cr}/\eta\Gamma_{\rm sh}^2$. 

\begin{figure}
\includegraphics[width=0.5\textwidth]{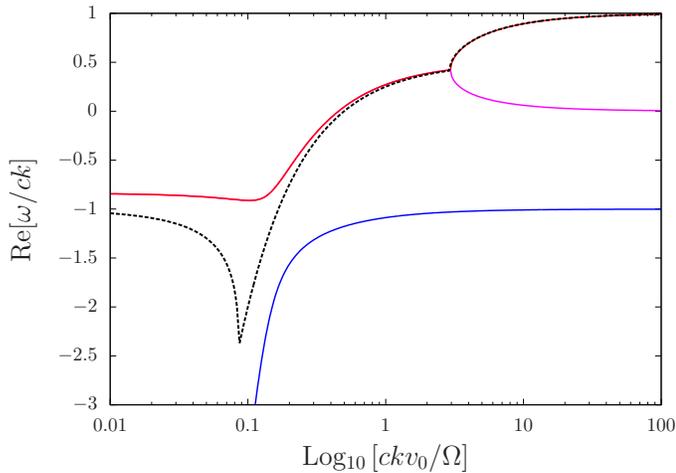} 
\caption{Real part of the dispersion relation (dashed line) from equation (\ref{Eq:OmegaPll}), for a parallel shock.
For comparison the solid lines are the solutions to the dispersion relation for the more detailed model given in the appendix.
The lines overlap in the range of interest. The shock Lorentz factor( measured in the proton frame) is 
$\Gamma_{\rm sh}=105$, $\eta=0.022$, and $\Omega^2/\omega_{\rm pp}^2=2\times10^{-5}$.}
\label{Fig:1}
\end{figure}

\begin{figure}
\includegraphics[width=0.5\textwidth]{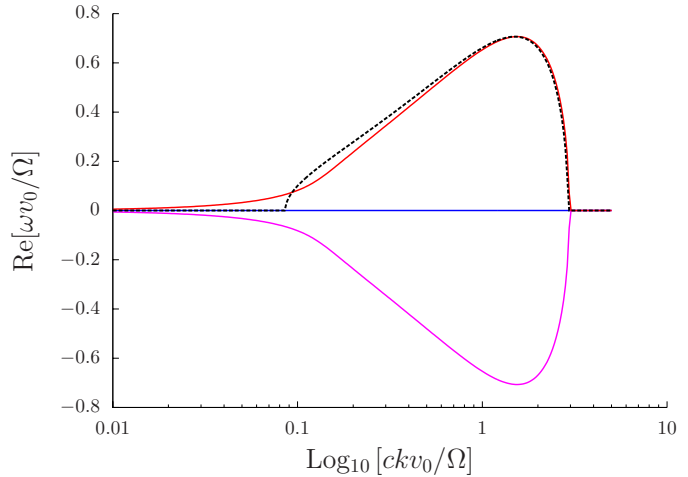} 
\caption{Comparison of the growth rates for a  parallel shock, given by Equation (\ref{Eq:OmegaPll}) (dashed line), with 
those found using a fully kinetic treatment (solid lines), for the same conditions as in Figure \ref{Fig:1}}
\label{Fig:2}
\end{figure}

Finally we note that this result, for exactly parallel shocks, can be reproduced using a more detailed kinetic analysis, which includes 
electron inertia, displacement current, finite temperature effects, as well as feedback on the cosmic rays. This result is
given in the appendix, following the approach of \cite{Revilleetal06}. Figures \ref{Fig:1} and \ref{Fig:2} show a comparison of the dispersion 
relation, as given in equation (\ref{Eq:OmegaPll}), with the numerical solutions of equation (A7). Both the real and imaginary parts of the dispersion 
relation are well approximated in the regime of interest, demonstrating that the assumptions made in the derivation of the dispersion relation are 
acceptable.

\subsection{Perpendicular shock}
\label{Sec:Perp}

The more pertinent case is that of a purely perpendicular shock. Certainly in the frame of the shock, the field is very close to 
perpendicular, due to the Lorentz transform of the perpendicular magnetic field component, 
unless it is closely aligned with the shock normal. As we have shown in the previous section, the perpendicular
component of the field is also compressed due to the drifting of the electrons to compensate the cosmic-ray current. Previous 
investigations of plasma instabilities operating at perpendicular shocks have worked in the limit of unmagnetised particles 
\cite[e.g.][]{LemoinePelletier10,Shaisetal12} which can not access the scales of interest  here.
\cite{Pelletieretal09} carried out a single-fluid MHD plus cosmic-ray analysis for magnetised shocks,
however, as shown in section 3, a single fluid MHD treatment is no longer viable in the ultra-relativistic limit.

We consider the shock normal again in the negative $z$-directions, with magnetic field in the $y$-direction. 
The acceleration of the electron frame in this case is $\bm{g} = cv_0 \Omega  \bm{\hat{x}}$, and 
all modes are measured in this non-inertial frame. Looking for plane wave solutions to equation (\ref{Eq:LinMom2}), we find
\eqb
\omega^2(\omega-ck_zv_0)^2-\omega^2\left(\frac{\Omega}{2\gamma_0}\right)^2-\left(\frac{\Omega}{2}ck_yv_0\right)^2=0
\enspace .
\eqe
Clearly in the limit of $k_y=0$ all modes are stable. Unstable solutions can be admitted however if $k_y \neq 0$.
We focus on $k=k_y$ modes, in which case calculating the dispersion relation is straightforward. Unstable modes are 
found for all $k$ in our long-wavelength approximation, although the 
instability will be suppressed at large $k$ when magnetic tension becomes important. At these scales, however, 
the particles are all unmagnetised, and the ion-Weibel instability will dominate \cite[e.g.][]{Shaisetal12}. The
instability is, in our approximation, almost purely growing at all wavelengths, and can be separated into two distinct regimes
\eqb
\label{Eq:RTgrowth}
{\rm Im}(\omega) = \left\lbrace\begin{array}{ccc}
 \gamma_0 ck_y v_0& &ckv_0 \ll \Omega/4\gamma_0^2 \\
\sqrt{\Omega ck_y v_0/2} & & ckv_0 \gg \Omega/4\gamma_0^2
\end{array}\right.
\eqe

The latter regime, being more rapid, is of greater interest here, and we focus on this instability in what follows.
It is easily demonstrated that the linearised equations, allowing only $k_y$ modes, can be reduced 
to the following set of coupled equations,
\eqb 
\label{Eq:CoupledEqns}
\pdiff{v_x}{t}= \frac{\Omega}{2}v_0\frac{n_1}{n_0} \enspace&,&  \pdiff{v_y}{t}=\Omega v_0 b_x \enspace ,\nonumber \\
\pdiff{}{t}\frac{n_1}{n_0}=-c\pdiff{v_y}{y} \enspace&,&  \pdiff{b_x}{t}=\frac{c}{2}\pdiff{v_x}{y}\enspace . 
\eqe
Motion in the $z$-direction can be neglected at short wavelengths.
Considering a linearly polarised wave $\delta B_x(y)$, it is readily seen that the thermal protons are deflected in the $y$-direction
by the Lorentz force, resulting in compressions and rarefactions along the $y$-axis. This bunching produces a perturbed current
$\delta j_z = e n_1 v_0$ which accelerates plasma in the $x$ direction. The resulting plasma motions have negative and 
positive regions of shear flow (vorticity) in regions of negative and positive $B_x$ respectively, which results in further growth 
of magnetic field, causing a runaway instability.

The similarity between this instability and the well known Raleigh-Taylor instability is self-evident. In particular, recalling the
acceleration of the non-inertial frame, $g= cv_0 \Omega$, the growth rate can be written in the more familiar form:
\eqb
\omega = {\rm i} \sqrt{\frac{gk}{2}}
\eqe

The instability for perpendicular shocks can thus be understood in terms of a Raleigh-Taylor instability, 
due to the effective gravity introduced by the zeroth order $\bm{\beta}_{0}\times\bm{B}_0$ force. The linear analysis
presented above, will clearly break down for $n_1>n_0$, however, further mixing/penetration of
high-density spikes, will allow continued 
non-linear growth of the field, should sufficient time be available. 

Having identified the relevant instabilities that can operate on 
long wavelengths at ultra-relativistic shocks, we now address the issue of whether sufficient time is available to allow significant 
field growth on the scales of interest, as this is most crucial when it comes to cosmic-ray acceleration.

\section{Application to cosmic-ray acceleration} 
\label{Sec:acc}

The acceleration of UHECRs requires growth of magnetic field fluctuations to a sufficient level
and most importantly on a sufficient scale. If this is to be seeded in the upstream plasma, the
instabilities described above, as a minimum requirement, must undergo at least one e-folding time.
As mentioned previously, the size of the precursor will depend sensitively on the particle 
scattering, since it is only necessary to deflect the particle through an angle $\sim 1/\Gamma_{\rm sh}$
(measured in the upstream frame), before the shock quickly overtakes it. We adopt a simple picture for the
details of energetic particle deflections in the shock precursor, a more detailed discussion of the particle scattering behaviour 
can be found in \cite{Achterbergetal} or \cite{Couchetal}. Note, however, that all calculations should be made in the frame in 
which the zero-th order electric field vanishes, which is different from previous authors.

\subsection{Parallel shocks}

As mentioned above, for a parallel (or indeed any subluminal) shock, in the absence of any pre-existing turbulence, 
the particles can outrun the shock indefinitely. However, if a small level of fluctuating magnetic field 
is introduced, $\delta B \ll B_0$, the particles can scatter/deflect on these perturbations, and 
eventually recross the shock. The fastest growing modes grow at wavelengths close to resonance 
with the thermal protons' Larmor radius, as measured in the electron drift frame, i.e. the protons have bulk Lorentz 
factor $\gamma_{\rm p} = \sqrt{2\eta}\Gamma_{\rm sh}$, and Larmor radius $\gamma_{\rm p}mc^2/eB_0$.
Note that $B_0$ is parallel to the shock normal, and hence invariant to boosts along this direction.
The cosmic-ray Larmor radius is expected to be orders of magnitude larger than that of the thermal protons, 
such that the growth of modes at resonant wavelengths with the cosmic rays
is too slow to have any significant effect, as is evident from inspection of Figure \ref{Fig:2}. Thus, resonant  
scattering is negligible unless non-linear effects can cause fields to grow to larger scales, as has been demonstrated to
occur in simulations of non-relativistic shocks \citep{RevilleBell13,Belletal13}. This requires on the order of 
5--10 e-folding times in the upstream plasma, before a fluid element is overtaken by the shock. 

We consider a simple pitch-angle diffusion 
model for the cosmic rays upstream of the shock, continuing to 
work in the electron drift frame. If the waves are initially on a scale $\lambda \ll r_{\rm g,cr}$, 
a crude approximation to the pitch angle diffusion coefficient is 
$$\mathcal{D}_\theta = \frac{<\Delta \theta ^2>}{2 \tau_{\rm sc}}\approx \left(\frac{\delta B_\bot}{B_0}\right)^2 \frac{c\lambda}{r_{\rm g,cr}^2}$$   
If the cosmic ray, entering the upstream with velocity $\beta_z(0)$ normal to the shock, undergoes a series of small angle deflections,
we can approximate the average deceleration as
\eqb
\beta_z(t) \approx \beta_z(0)\left[1-\frac{1}{2}\bm{\vartheta}^2\right]  \sim \beta_z(0)\left[1-\mathcal{D}_\theta^2 t\right]  
\eqe
The maximum precursor size is determined from the condition $\beta_z(t)-\beta_{\rm sh,e}=0$, 
which in the limit $\beta_z(0) = 1$ is
$$L_{\rm pre} \approx \frac{r_{\rm g,cr}^2}{\lambda \gamma_{\rm sh}^4}\left(\frac{B_0}{\delta B_\bot}\right)^2$$
The corresponding maximum growth time for streaming instabilities to develop is $t_{\rm grow}=L_{\rm pre}/\beta_{\rm sh,e} c$.

Inserting the maximum growth rate from equation (\ref{Eq:PllMax}), with $\lambda = 2\pi/k_{\rm max}$, using the values 
for $\gamma_{\rm p},~\gamma_{\rm sh}$ and $\gamma_{\rm cr}$ given in 
section \ref{Sec:Param}, it follows that
\eqb
\omega_{\rm max} t_{\rm grow} &\approx&
\frac{3}{4\pi}\frac{\gamma_{\rm cr}^2}{\gamma_{\rm p}^2\gamma_{\rm sh,e}^4}\left(\frac{B_0}{\delta B_\bot}\right)^2
\nonumber\\
&\approx&(1-8\eta)^{-1/2}\left(\frac{B_0}{\delta B_\bot}\right)^2
\eqe
which, for $\eta \ll 1$, is essentially independent of the parameters of the reflected component. For field growth to  
extend to larger scales, this number must exceed unity. However, the growth of magnetic field on small 
scales appears to self-limit the growth of the instability, since the number of e-foldings decreases rapidly as 
$\delta B$ increases, implying that magnetic field amplification is limited to $\delta B \lesssim B_0$ and to scales $\ll r_{\rm g,cr}$. 

We again note that particles that cross the shock multiple times, with higher energies, can extend further into the upstream region,
and allow for larger growth. As mentioned previously, the fraction of the energy $\eta$ carried by these particles 
will be sensitive to the details of the particle spectrum. 
There are suggestions from Monte Carlo or PIC simulations, that a non-thermal tail $dn/d\gamma \propto \gamma^{-2.2}$ is produced
\citep{Achterbergetal,SironiSpitkovsky11}, although other report considerably steeper spectra \citep{Ostrowski,SummerlinBaring}. As the energy
density in the beam decreases, the growth rate reduces to that found in \cite{Revilleetal06}, which can have a larger maximum growth-rate,
but at increasingly smaller length scales. 

Parallel shocks represent a singular case, since the problems of maximum energy associated with the mean 
field transporting the highest energy particles into the downstream do not arise. While this does not 
rule out the possibility of multiple shock crossings, and acceleration to higher energies, it appears impossible to 
enter a regime where resonant scattering can occur at high energies.  The problem thus becomes a time limitation 
at high energies. The isotropisation time $\mathcal{D}_{\theta}^{-1}$, for UHECRs will become longer than the age of any realistic system.
Similar conclusions were drawn by \cite{sironietal13}, for acceleration in purely small-scale fields. Adopting the Blandford-McKee solution 
for a relativistic expanding blast-wave \citep{BlandfordMcKee76}, they found a maximum energy $\sim 10^{17}$~eV, in the ISM frame.

\subsection{Perpendicular shocks}

For perpendicular shocks, the situation is more straightforward, as to lowest order, particles simply undergo a regular deflection in the upstream field,
as measured in the electron frame i.e. the frame in which the zeroth order electric field vanishes. 
Following the same argument as above, the precursor length is 
\eqb
L_{\rm pre} = \frac{r_{\rm g,cr}}{\gamma_{\rm sh}^3}
\eqe
and maximum growth time as before. Using the shorter wavelength solution in (\ref{Eq:RTgrowth}) 
the number of e-foldings at a given wavenumber is
\eqb
\omega t_{\rm grow} \approx \sqrt{\frac{\Omega}{2}ck_yv_0}\frac{r_{\rm g,cr}}{c\gamma_{\rm sh}^3}\enspace .
\eqe
Since the growth at longer wavelengths is slower, this can be considered as an upper limit to the growth 
in general. Since the number of e-foldings increases indefinitely with wavenumber, $k$, it suffices to determine
the wavelength at which one e-folding is achieved. This occurs at
\eqb
 k r_{\rm g,cr} = \frac{2}{v_0}\gamma_{\rm sh}^6\frac{\gamma_0}{\gamma_{\rm cr}}\approx
\frac{1}{(8\eta)^2}\frac{\Gamma_{\rm sh}^2}{\Gamma_{\rm cr}} \enspace,
\eqe
where, as usual, lower case $\gamma$s refer to quantities measured in the electron rest frame. At the injection energy 
$\Gamma_{\rm cr} \sim \Gamma_{\rm sh}^2$, clearly $ k r_{\rm g,cr} \gg 1$, i.e. non-linear growth can 
only take place on wavelengths much less than gyro-radius of the cosmic rays. At higher energies,
$\Gamma_{\rm cr} \gg \Gamma_{\rm sh}^2$, this value will depend on the shape of the spectrum.
However, most results to date put the shape of the spectrum to be steeper
than $dn/d\gamma \propto \gamma_{\rm cr}^{-2.2}$. Thus $\eta\propto \gamma_{\rm cr}^{-1.2}$ or steeper,
such that the scales that one can expect non-linear growth, and the scale of the Larmor radius diverge with increasing energy. We note 
that since $k$ lies along the field, and is linearly polarised in the plane orthogonal to the shock normal, and
magnetic field, the length and relative magnitude, with respect to the mean field, are preserved. Since it is scattering in the
downstream that determines whether a particle is advected downstream or not, it is the gyroradius of the cosmic ray in
this frame that is most important. The gyro radius of a cosmic ray with Lorentz factor 
$\gamma_{\rm cr}$ in the electron frame has, on crossing into the downstream, a Larmor radius 
\eqb
r_{\rm g,d} = \sqrt{8\eta}\gamma_{\rm cr}\frac{m c^2}{e \bar{B}_{\rm d}}={8\eta}\gamma_{\rm d}\frac{m c^2}{e \bar{B}_{\rm d}}\enspace,
\eqe
where $\bar{B}_{\rm d}$ is the shock compressed mean field. It follows that 
\eqb
\label{krglimit}
 k r_{\rm g,d} \approx
\frac{1}{8\eta}\frac{\Gamma_{\rm sh}^2}{\Gamma_{\rm cr}} 
\eqe
which diverges less quickly, but still exceeds unity. Since it is not possible to achieve even one e-folding close to 
gyro-resonance, the scattering waves for high-energy particles, close to the values given in equation (\ref{Eq:gammaxd}) 
are not seeded upstream, via the plasma instabilities found in this paper. It is interesting to note that it is still possible to have 
several e-foldings on shorter wavelengths, which may allow for more rapid acceleration of \lq lower\rq~energy electrons (lower than 
the equations (\ref{Eq:gammaxd}) or (\ref{Eq:EMAXXX}) limit).

Perpendicular shocks also offer another possibility for growth, due to either large-scale 
fluctuations pre-existing in the ambient medium, or non-uniform injection of particles over the shock
surface. The former case will preserve the relative amplitude of the mean field with the perpendicular 
fluctuations, and as such, it has a negligible effect by itself, unless the magnetic field threading the 
ambient medium is highly non-uniform $\left|\bf{\delta B}\right| \gg \langle \bm{B}\rangle$, e.g. for 
interaction with a striped wind. This would likely introduce a characteristic length/energy scale, which 
is disfavoured by the observations, which typically have power-law form \citep{Bandetal93}. 

The latter case, of non-uniform injection over the shock surface can be 
treated in a reduced model. It is clear from the analysis 
in section \ref{Sec:Perp}, the return current is provided by the drift of the background protons in the electron
frame, which previously was used to determine the acceleration of the non-inertial frame 
\eqb
\frac{ d \bm{p}_{\rm p}}{dt} = e \bm{\beta}_0\times\bm{B}_0
\eqe 
where $B_{0}$ is the compressed value measured in the electron drift frame, described in section \ref{Sec:Param}.
If the injection is non-uniform, $\bm{B}_0$ varies in the plane of the shock, and
neighbouring fluid elements will be differentially accelerated. The field will therefore shear, leading to amplification.
Assuming the perpendicular velocity does not become relativistic, the displacement of neighbouring field points is 
\eqb
\frac{d^2 \xi}{d t^2} \approx \frac{eB_\bot(\bm{r})}{\gamma_{\rm p}m_{\rm p}} \beta_0\enspace.
\eqe
The maximum possible separation of neighbouring fluid element in the precursor can be estimated assuming neighbouring 
regions are at the two extremes $\eta_1\Gamma_{\rm sh}\gg 1$, and $\eta_2\Gamma_{\rm sh} \approx 0$, in which case
\eqbn
\Delta \xi_{\rm max} < \frac{\gamma_{\rm cr}}{\gamma_{\rm p}\gamma_{\rm sh}^6}r_{\rm g,cr}
\eqen
where all terms are again measured in the electron drift frame. Since this perturbation is perpendicular to both the 
shock normal and mean field, it is preserved on crossing the shock. Using the values from section \ref{Sec:Param},
and the above expression for $r_{\rm g,d}$, it follows that 
\eqb
\frac{\Delta \xi_{\rm max} }{r_{\rm g,d}} < \eta \frac{\Gamma_{\rm cr}}{\Gamma_{\rm sh}^2} \enspace,
\eqe
which, modulo a factor of order unity, is essentially equivalent to the equation (\ref{krglimit}), which has already been demonstrated to be insufficient.

\subsection{Instabilities downstream of the shock}

The above instabilities, occurring in the foreshock region, are insufficient to provide effective 
scattering of high energy particles.
It is possible, however, that currents downstream of the shock can excite plasma instabilities. It is well
known that the distribution of cosmic rays is highly anisotropic at relativistic shocks. As measured in the downstream frame, 
the distribution function is peaked at pitch angles closely aligned with the plane of the shock \citep{Kirketal00}.

For all but a small fraction ($\sim 1/\Gamma_{\rm sh}$) of possible field configurations, the magnetic field downstream of the shock
lies in the plane of the shock. The anisotropic distribution thus produces a current, which we can take to orthogonal to both the shock normal 
and the magnetic field. The situation is therefore similar to the perpendicular field case 
considered above, only now the background plasma is relativistically hot, with mean temperature $\Gamma_{\rm sh} m c^2$.
The magnitude of the cosmic-ray current will depend on the degree of anisotropy, but for solutions given in \cite{Kirketal00}, 
immediately downstream of the shock it is a sizeable fraction of the speed of light ($\sim0.5$).
This current will persist in the downstream until it has been isotropism, which is again determined by the diffusion coefficient  $D_{\theta}$.
Assuming scattering is dominated by deflections in small-scale Weibel structures, the current exists in the downstream frame for a time
\eqb
t_{\rm grow} \sim D_\theta^{-1} 
\eqe

To calculate the other relevant parameters, we assume the cosmic rays are overtaken by the shock without
significant scattering and with pitch angle $\theta\gtrsim 1/\Gamma_{\rm sh}$.
Making a Lorentz transform into the downstream frame results in a reduction by a factor $\sim \Gamma_{\rm sh}$ 
of both the average energy per particle as well as the number density of the cosmic-rays.
The background protons, on the other hand, are compressed, with proper compression ratio $\sqrt{8}\Gamma_{\rm sh}$. 
The total number density of cosmic rays in the upstream  is determined by particles at the injection energy $\Gamma_{\rm sh}^2 mc^2$, 
such that in the downstream $\bar{n}_{\rm cr}\sim \eta \bar{n}_{\rm p}$, and decreasing as before for higher energies.

If the electrons are also thermalised in the downstream proton frame, the background plasma can be treated as a single fluid, 
and we can take as an upper limit on the acceleration of a Lagrangian fluid element, with all quantities are measured in the downstream frame,
\eqb
\label{Eq:dsgrowth}
\frac{d^2{\xi}}{dt^2} < \frac{e n_{\rm cr}\beta_{\rm drift} \langle B_{\rm d}\rangle }{\Gamma_{\rm sh} m_{\rm p} n_{\rm p}}\sim \eta \beta_{\rm drift} 
\frac{e \langle B_{\rm d}\rangle }{\Gamma_{\rm sh} m_{\rm p}}\enspace.
\eqe  
where $ \langle B_{\rm d}\rangle$ is the shock compressed mean field $\sim \sqrt{8}\Gamma_{\rm sh} B_\bot^0$.
 
 Using the expression for $D_\theta$ given in equation (\ref{Eq:Dtheta}), and taking the parameters on the right hand side of (\ref{Eq:dsgrowth})
 not to vary in time, the maximum displacement is 
 \eqb
 \frac{\Delta \xi_{\rm max}}{r_{\rm g,d}}< \frac{\eta\beta_{\rm drift}}{{\Gamma}^2_{\rm sh}} \frac{\sigma_u}{\sigma_d} \left(\frac{\gamma_{\rm d}}{{\Gamma_{\rm sh}}}\right)^3
\left(\frac{c/\omega_{\rm pp}}{\lambda}\right)^2=\frac{\eta\beta_{\rm drift}}{\gamma^{2}_{\rm d,max}}  \left(\frac{\gamma_{\rm d}}{{\Gamma_{\rm sh}}}\right)^3 \nonumber
\eqe
 where $r_{\rm g,d}$ is the cosmic-ray gyro radius measured in the mean field, as in the previous section,
 and $\gamma_{\rm d, max}$ is as given in equation (\ref{Eq:gammaxd}). Again, this 
 is insufficient at low energies $\gamma_{\rm d}\sim \Gamma_{\rm sh}$, and does not improve at higher energies, on account of $\eta$.


\section{Discussion}

Ultra-relativistic shocks are frequently identified as, or associated with strong sources of non-thermal radiation. Shock acceleration
offers a well-tested mechanism for generating the non-thermal particle populations required to produce this
emission, and theory and simulations are converging to provide a more complete picture of the plasma processes occurring in these
environments. From a  phenomenological perspective, observations can typically be explained using simple leptonic models,
whereas extracting information about non-thermal hadronic populations is challenging. However, on theoretical grounds, 
it is expected that any source capable of accelerating protons or nuclei to ultra-high energies $> 10^{18}$~eV, is likely also to 
accelerate electrons rapidly, at least to their radiation reaction limit. This motivates consideration of strong gamma-ray emitters, 
such as GRB, pulsars and AGN as potential candidates for UHECR production, all of which are thought to contain 
ultra-relativistic shocks.

Motivated by recent successes in numerical modelling of the micro-physics of relativistic shocks, 
the acceleration of particles at weakly-magnetised ultra-relativistic shocks has been investigated.
We demonstrate that Weibel-mediated shocks, or indeed any shock mediated by kinetic instabilities, 
operating at the plasma-skin depth, cannot accelerate particles above a critical energy, given in equation 
(\ref{Eq:EMAXXX}) \cite[see also][]{LemoinePelletier10}: 
\eqb E_{\rm max} \approx
\left(\frac{\Gamma_{\rm sh}}{100}\right)^2 \left(\frac{\lambda_{\rm d}}{10c/\omega_{\rm pp}}\right)
\left(\frac{\sigma_{\rm d}}{10^{-2}}\right)\left(\frac{\sigma_{\rm u}}{10^{-8}}\right)^{-1/2}~\mbox{PeV~,} \nonumber
\eqe
Acceleration to higher energies can only be achieved if strong scattering can occur on scales $\gg c/\omega_{\rm pp}$.
We have investigated the growth of plasma instabilities, driven by cosmic-ray currents, both upstream and downstream of
an ultra-relativistic shock in this long-wavelength limit. In all cases, it is demonstrated that the growth of any such
instability is too slow, on the scales required to facilitate acceleration to higher energies. As such, our results suggest that the above 
energy limit can not be circumvented, implying this maximum energy is an inherent limitation of shock-acceleration at
ultra-relativistic shocks. Future simulations that can self-consistently investigate the scattering of ultra-relativistic particles in 
self-generated relativistic-plasma turbulence may ultimately be required to provide confirmation.

We emphasise that it is not suggested that the instabilities presented in this paper do not play any role. In fact, 
they may still lead to 
significant growth of magnetic field on scales larger than the ion collisionless skin depth, which may increase 
the rate of acceleration for lower energy particles. This may be an essential feature in the case of electron acceleration
at GRB shocks, in the presence of radiation reaction limited acceleration \citep{KirkReville}. 
 
While this paper is by no means the first to suggest that GRBs are not the source of UHECRs
\cite[e.g.][]{MilosNakar06}, we have gone a 
step further,  in demonstrating that ultra-relativistic shocks are dis-favoured as sources of high energy particles in general. 
The maximum energy, given above, for typical parameters expected in external GRB shocks, for example,  
may have a detectable signature, which is measurable with the next generation Cherenkov 
observatory, CTA \citep{CTA}. 

Processes other than shock acceleration may provide additional mechanisms for UHECR production in
ultra-relativistic jets \cite[e.g.][]{Ostrowski98,RiegerDuffy}, or around rapidly rotating compact objects 
 \cite[e.g.][]{Bell92,Fangetal13}.
However, the acceleration mechanisms operating in such flows are not near as well established as the Fermi
shock acceleration process, although this may change with time. Diffusive
shock acceleration at large-scale non-relativistic shocks, such as those expected to be found in galaxy mergers, 
or AGN radio \lq hot-spots\rq~, for example, offer alternative, highly plausible candidates.

\section*{Acknowledgments}

The research leading to these results has received funding from the European Research Council under the 
European Community's Seventh Framework Programme (FP7/2007-2013) / ERC grant agreement no. 247039.


 \appendix

\section{Kinetic treatment for parallel shock case}

The dispersion relation for circularly polarised waves propagating parallel to the mean field is \cite[e.g.][]{Revilleetal06}
\begin{eqnarray}
\frac{k_\|^2c^2}{\omega^2}-1&=&\sum_j \chi_j
\end{eqnarray}
where
\eqb
\chi_j=\Gamma_j
\frac{\omega_{{\rm p}j}^2}{\omega^2}\int\frac{\diff^3 u}{\gamma} f_j(\vec{u})
\left[\frac{-\omega\gamma+cku_\|}{D\left(u_\|\right)} 
- {u_\bot^2\over 2}{(c^2k^2-\omega^2)\over D^2\left(u_\|\right)}\right] \enspace \nonumber
\label{yoon}
\eqe
is the susceptibility for each plasma component. We work with the normalised 4-momentum $u^\mu = dx^\mu/ds$, 
and $\omega^2_{{\rm p}j} = 4\pi n_j q_j^2/\Gamma_jm_j$ and $\omega_{{\rm c}j}=q_jB_0/m_jc$
are the relativistic plasma frequency and gyro-frequency for each species $j$. Here,
$n_j,\Gamma_j=\int (1,\gamma) f_j({\bm u}) d^3u$ is the mean density, Lorentz factor of each species.
The resonant denominator is
\eqb
D\left(u_\|\right)&=&\varepsilon\omega_{{\rm c}j}\left(1+
Z\left(u_\|\right)\right)
\label{resdenom}
\\
\noalign{\hbox{with}}
\nonumber\\
Z\left(u_\|\right)
&=&{\omega\gamma-cku_\|\over\varepsilon\omega_{{\rm c}j}}
\label{zsmall}
\eqe
and the waves have left(right)-handed polarisation for $\varepsilon=+1(-1)$, for $k>0$.
It is readily noticed that $\omega^2\chi_j$ is an invariant quantity, allowing us to calculate the 
dispersion relation in an arbitrary frame. 
 
We consider a three component plasma, protons, electrons and cosmic-rays (also protons),
all components being treated as cold beams
\eqb 
f_{j}(u_\|,u_\bot) = \frac{1}{2\pi u_\bot}\delta(u_\bot)\delta(u_\| - \Gamma_{j}\beta_{j})
\eqe
The susceptibility of each component is thus
\eqb
\omega^2\chi_{j}(k,\omega)&=&
{ \omega_{{\rm p}j}^2\Gamma_j\left(c\beta_{j}k-\omega\right)
\over
\varepsilon\omega_{\rm cj} - \Gamma_{\rm j}\left(c\beta_{j}k-\omega\right)}
\eqe

Assuming the electrons are magnetised in their own rest frame,  
$\Gamma_{\rm e}\left|\omega- c\beta_{\rm e}k\right|\ll\left|\omega_{\rm ce}\right|$, 
the denominator of the electron susceptibility can be expanded to give, 
\eqb
\omega^2\chi_{\rm e}=\Gamma_{\rm e}
\frac{\omega_{\rm pe}^2}{\varepsilon\omega_{\rm ce}}
\left(c\beta_{\rm e}k-\omega\right)+\Gamma_{\rm e}^2
\frac{\omega_{\rm pe}^2}{\omega_{\rm ce}^2}
\left(c\beta_{\rm e}k-\omega\right)^2 + \dots
\eqe

Working from this point on exclusively in the electron rest frame ($\beta_{\rm e}=0$), 
using subscript b to represent the cosmic-ray protons, the full dispersion relation reads 
\eqb
\label{Eq:A1}
&&k^2c^2-\left(1+\frac{\omega_{\rm pe}^2}{\omega_{\rm ce}^2}\right)\omega^2 +\frac{\omega_{\rm pe}^2}{\varepsilon\omega_{\rm ce}}
\omega+\\
&&~~~\frac{ \omega_{\rm pb}^2\Gamma_{\rm b}\left(\omega-c\beta_{\rm b}k\right)}{
\varepsilon\omega_{\rm cb} + \Gamma_{\rm b}
\left(\omega-c\beta_{\rm b}k\right)
}+
\frac{ \omega_{\rm pp}^2\Gamma_{\rm p} \left(\omega-c\beta_{\rm p}k\right)}
{\varepsilon\omega_{\rm cp} + \Gamma_{\rm p}
\left(\omega-c\beta_{\rm p}k\right)
}
= 0 \nonumber
\eqe
The roots of this equation are found numerically and are plotted in Figures \ref{Fig:1} and \ref{Fig:2}. 

The dispersion relation, equation (\ref{Eq:OmegaPll})
can also be derived from this equation, and helps demonstrate the limitations of the approximate approach
used in the main section of the paper. Neglecting the first three terms on the left hand side of 
equation (\ref{Eq:A1}), and taking the ultra-relativistic limit, $n_{\rm e}=2n_{\rm p}=2n_{\rm b}$, in
the limit of $\beta_{\rm b} = -1$, the dispersion relation for  $\varepsilon=+1$ can be expressed as
\eqb
\label{Eq:A8}
\tilde{\omega}^3+\frac{1+z}{2}\tilde{\omega}^2+\tilde{k}\left[
1-\beta_{\rm p}(z+\tilde{k})\right]\tilde{\omega}
+ \frac{1+z}{2}\beta_{\rm p} \tilde{k}^2 = 0
\eqe
where $z=\Gamma_{\rm p}/\Gamma_{\rm b}\ll1$, and 
$$\tilde{\omega} = \Gamma_{\rm p}\omega/\omega_{\rm c}
\mbox{~~~~~~~~~~~~~and~~~~~~~~~~~~}   
\tilde{k} = \Gamma_{\rm p}kc/\omega_{\rm c}\enspace .$$
In deriving (\ref{Eq:A8}), all terms containing $1-\beta_{\rm p}$ have been neglected. An approximate solution to this cubic can be found
by solving for $\tilde{k}$,
\eqb
\frac{\beta_{\rm p}k}{\omega} = \frac{1-\beta_{\rm p}z\pm2\beta_{\rm p}^{1/2}\sqrt{\omega^2-z}}{2\omega-(1+z)}
\eqe
where we have dropped the tildes. 
For modes with $\left|\omega^2\right|\gg z$, the frequency for unstable modes is
\eqb
&&\omega = \frac{1}{2}\left(\beta_{\rm p}k-\frac{1-z\beta_{\rm p}}{2}\right)\pm \\
&&~~~~~~~~\frac{1}{2}
\sqrt{\left(\beta_{\rm p}k-\frac{1-z\beta_{\rm p}}{2}\right)^2+2z-2(1+z)\beta_{\rm p}k}\nonumber
\eqe
which clearly reduces to (\ref{Eq:OmegaPll}) in the limit $z\rightarrow0$.
\label{lastpage}

\end{document}